\documentclass{article}

    \PassOptionsToPackage{numbers, compress}{natbib}

\usepackage[final]{neurips_2023_ml4ps}  

\usepackage[utf8]{inputenc} 
\usepackage[T1]{fontenc}    
\usepackage{url}            
\usepackage{booktabs}       
\usepackage{amsfonts}       
\usepackage{nicefrac}       
\usepackage{microtype}      
\usepackage{xcolor}         
\usepackage{amsmath}
\usepackage{mathtools}
\usepackage{braket}
\usepackage{natbib}
\usepackage{graphicx}
\usepackage{cite}
\usepackage{ftnxtra}
\usepackage{fnpos} 
\usepackage{subcaption}
\usepackage{hyperref}

\DeclareMathOperator{\EX}{\mathbb{E}}

\DeclareMathOperator{\vx}{\boldsymbol{x}}

\DeclareMathOperator{\Ne}{N_{e}}
\DeclareMathOperator{\rhom}{\rho_{{\cal M}}}
\DeclareMathOperator{\rhozero}{\rho_{0}}
\DeclareMathOperator{\T}{T(\cdot)}

\DeclareMathOperator{\vz}{\mathbf{z}}
\DeclareMathOperator{\hrhom}{\hat{\rho}_{{\cal M}}}

\title{Orbital-Free Density Functional Theory with Continuous Normalizing Flows}

\author{%
  Alexandre de Camargo\\
  Department of Chemistry and Chemical Biology\\
  McMaster University\\
\And
  Ricky T. Q. Chen \\
  FAIR, Meta \\
\And
  Rodrigo A.~Vargas-Hernández\\
  Department of Chemistry and Chemical Biology\\
  McMaster University\\
  \texttt{vargashr@mcmaster.ca} \\
}

\begin{document}

\maketitle

\begin{abstract}
Orbital-free density functional theory (OF-DFT) provides an alternative approach for calculating the molecular electronic energy, relying solely on the electron density. In OF-DFT, both the ground-state density is optimized variationally to minimize the total energy functional while satisfying the normalization constraint. In this work, we introduce a novel approach by parameterizing the electronic density with a normalizing flow ansatz, which is also optimized by minimizing the total energy functional. Our model successfully replicates the electronic density for a diverse range of chemical systems, including a one-dimensional diatomic molecule, specifically Lithium hydride with varying interatomic distances, as well as comprehensive simulations of hydrogen and water molecules, all conducted in Cartesian space.
\end{abstract}

\section{Introduction}
\vspace{-0.2cm}
Unlike Kohn-Sham density functional theory (KS-DFT), which relies on molecular orbitals~\citep{kohn1965_self_DFT}, Orbital-free density functional theory (OF-DFT) offers a distinctive computational approach within quantum chemistry and condensed matter physics.

\begin{figure}[!h]
    \centering
    \begin{subfigure}[t]{0.05\linewidth}
        \rotatebox[origin=l]{90}{\hspace{2.9em} \Large \texttt{H}$_2$}
    \end{subfigure}%
    \begin{subfigure}[t]{0.73\linewidth}
        \includegraphics[width=\linewidth]{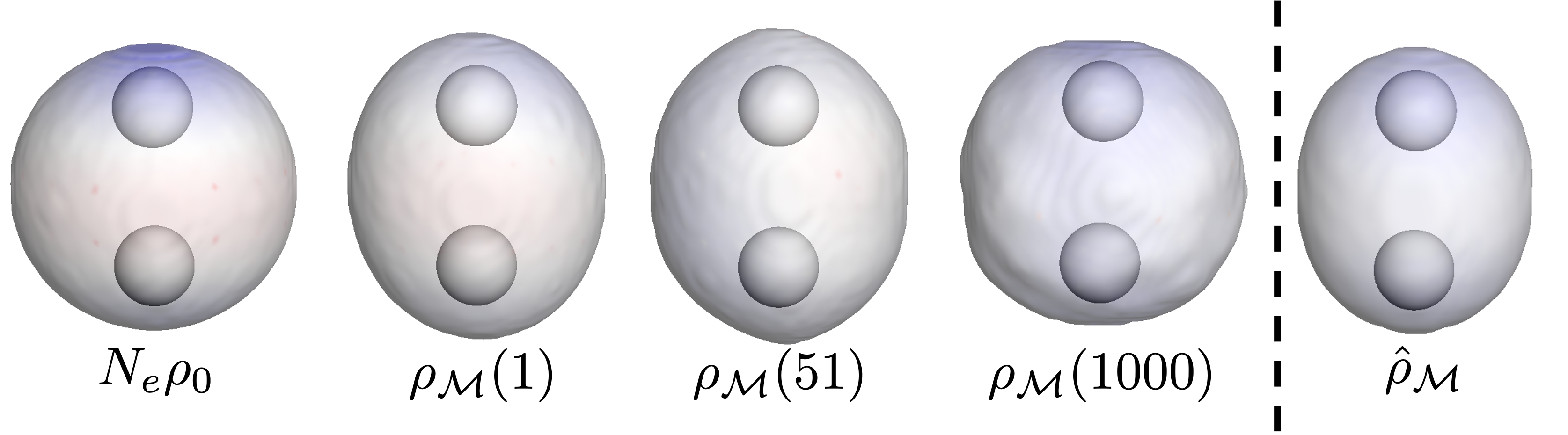}
    \end{subfigure}\\
    \begin{subfigure}[t]{0.05\linewidth}
        \rotatebox[origin=l]{90}{\hspace{2.4em} \Large \texttt{H}$_2$\texttt{O}}
    \end{subfigure}%
    \begin{subfigure}[t]{0.73\linewidth}
        \includegraphics[width=\linewidth]{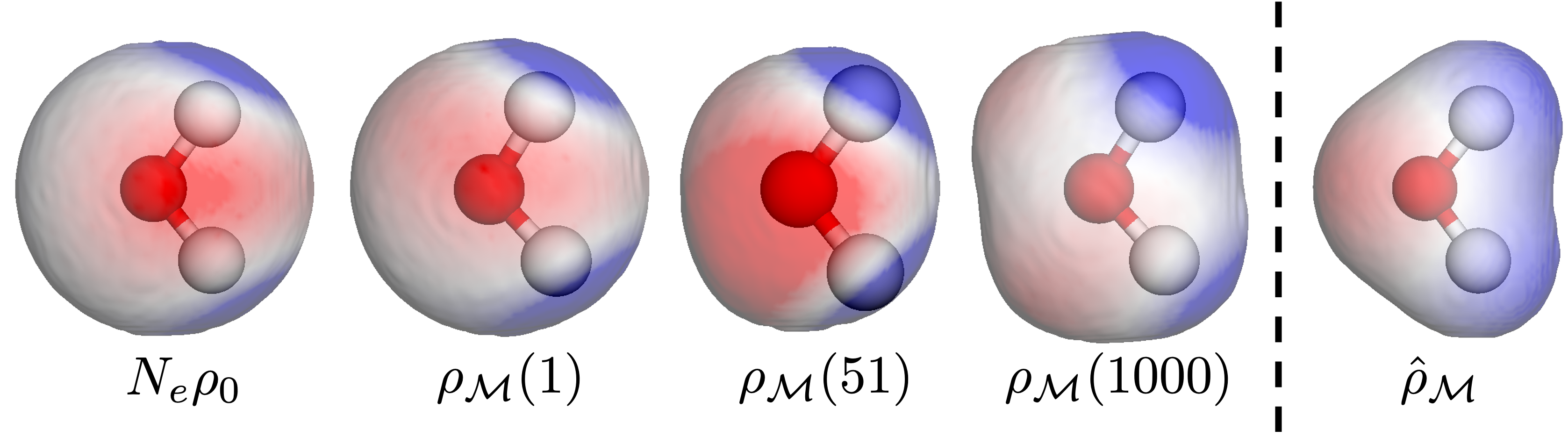}
    \end{subfigure}
    \caption{MEP for $\rhom(i)$ computed at different iterations $i$. $\hrhom$ and $\rhozero$ are defined in the main text. }
    \label{fig:MEP}
\end{figure}

It finds its roots in the Hohenberg-Kohn theorems~\citep{hohenberg1964_inhomogeneous,gpaar1980_DFT}, which unequivocally establish that the ground-state properties of a many-electron system can be ascertained through the minimization of an energy functional, denoted as $E[\rho]$, that operates solely on the electron density, $\rho$. This approach renders OF-DFT especially advantageous when grappling with vast and intricate systems, making it a valuable tool with wide-ranging applications in materials science~\citep{witt2018orbital}.

For a system with $N$ electrons subjected to an external potential $(v_{ext})$, the total energy functional is defined as $E[\rho] = F[\rho] + \int v_{\text{ext}}(\vx)\rho(\vx)\mathrm{d}\vx$ \citep{hohenberg1964_inhomogeneous,gpaar1980_DFT}. $F[\rho]$ represents the universal functional, with its precise form remaining unknown. Common approximations rely on kinetic ($T[\rho]$) and potential ($V[\rho]$) energy contributions, the specific forms of which also remain elusive. Numerous proposed functionals aim to approximate $T[\rho]$~\citep{witt2018orbital, kumar2022accurate,  bach2014many, carter2008challenges}, and machine learning (ML) algorithms have been employed for this purpose~\citep{snyder2012finding, brockherde2017bypassing, meyer2020machine, imoto2021order, pederson2022machine}. In contrast, within the context of KS-DFT, $T[\rho]$ is approximated as the sum of non-interacting particles~\citep{kohn1965_self_DFT,gpaar1980_DFT}. Researchers are continuously refining OF-DFT methods to enhance accuracy and broaden their applicability.


In OF-DFT, the ground-state density is found by solving a constrained optimization problem,
\begin{eqnarray}
    \min_{\rho(\mathbf{\vx})} E[\rho(\mathbf{\vx})] - \mu \left(\int \rho(\mathbf{\vx}) \mathrm{d} \mathbf{\vx} - \Ne \right )\label{ofdft_lagrangian} \ \text{s.t. } \rho(\mathbf{\vx}) \geq 0,
    \label{eqn:ofdft_min}
\end{eqnarray} 
where $\mu$, also known as the chemical potential, acts as the Lagrange multiplier associated with the normalization constraint on the total number of particles ($\Ne$). These constraints, which enforce both positivity and normalization, ensure the attainment of physically valid solutions.
Typically, conventional methods for solving for $\rho$ involve self-consistent procedures based on functional derivatives, leading to the Euler equation $\delta E[\rho(\vx)]/\delta \rho(\vx) - \mu = 0$~\citep{gpaar1980_DFT}.

To represent molecular densities in real space, grid-based models are commonly employed \citep{mortensen2005real, becke1988multicenter, wagner2012reference_1d_soft_coulomb, jiang2003density, bu2023efficient}. An alternative real-space formulation was proposed by \citet{chan2001thomas}, which involves expressing the density as a linear combination of Gaussian basis functions ($\eta_i$), denoted as $\varphi(\vx) = \sum_i C_i \eta_i(\vx)$, and ensuring positivity through $\rhom(\vx) = \varphi^2(\vx)$. Both approaches, mesh and basis set, do not guarantee the normalization constraint; however, the latter method assists in the computation of essential integrals required for estimating functional derivatives and other relevant quantities.

In the present work, we present an alternative constrain-free approach to solve for the ground-state density using normalizing flows (NF). \vspace{-10pt} 

\section{Proposed Method}
\vspace{-.2cm}
Our method consists of parametrizing the electronic density $\rho(\vx)$ with a Normalizing Flow. NFs are probabilistic models capable of transforming a simple density ($\rhozero$) into a potentially more complex target distribution, using the change of variables formula, \vspace{-.2cm}
\begin{equation}
    \rho_\phi(\mathbf{x}) = \rhozero(\mathbf{z})\left|\text{det}  \nabla_{\vz} T_\phi(\vz) \right|^{-1},
    \label{eqn:density_transform}
\end{equation}
where $T_\phi(\cdot)$\footnote[1]{$T_{\phi}: \mathbb{R}^{D} \xrightarrow{} \mathbb{R}^{D}$ an invertible and differentiable transformation. } is a bijective transformation, and $\rhozero(\vz)$ is the base distribution of the flow-based model. Eq. \eqref{eqn:density_transform}
guarantees the preservation of volume in the density transformation. 
To have a valid flow, we must have access to samples from $\rhozero$ and its probability density function (PDF) should be differentiable for optimization. 
A common approach is to parametrize $\T$ through a composition of functions; $T_\phi(\cdot) = T_{K}(\cdot) \circ \cdots \circ T_{1}(\cdot)$ ~\citep{papamakarios2021normalizing,rezende2015FineteFlow,kobyzev2020normalizing_review}. 
%
These composable transformations can be considered as a flow discretized over time. 
A continuous-time formulation was proposed by Chen et al. \citep{chen2018neuralODe}, referred to as continuous normalizing flows (CNF).
This framework is centered around the computation of the log density and $\T$ through an ordinary differential equation (ODE),\looseness=-1
\begin{eqnarray}
    \tfrac{\mathrm{d}}{\mathrm{d} t} \mathbf{z}(t) = g_{\phi}(\mathbf{z}(t),t), \text{ and  }     
    \tfrac{\mathrm{d} }{\mathrm{d} t} \log \rho(\mathbf{z}(t))= -
    \nabla \cdot g_{\phi},
    \label{eqn:cnf_evol}
\end{eqnarray}
where ``$\nabla \cdot$'' denotes the divergence operator.
In this work, $g_{\phi}(\cdot)$ is a neural network composed of three hidden layers each with 512 neurons and the hyperbolic tangent activation function. Other architectures were tested but found to be sub-optimal. 

Here, we reframe the OF-DFT variational problem as a Lagrangian-free optimization problem for molecular densities in the real space by parameterizing $\rhom(\vx)$ with a CNF ($\rho_\phi(\vx)$). 
Since the molecular density normalizes to $\Ne$, we define $\rhom(\vx)$ as, $\rhom(\vx) := \Ne \rho_\phi(\vx)$, guaranteeing the satisfaction of the normalization constrain. $\rho_\phi(\vx)$ is also known as \emph{shape factor} ~\citep{gpaar1980_DFT,gparr1983shapefactor}.

The value of any functional that conforms to the total energy functional ($E[\rhom]$) can be rewritten in terms of an expectation over $\rho_{\phi}$, \vspace{-.3cm}
\begin{eqnarray}
    F[\rhom] = \int f(\vx,\rhom,\nabla\rhom)\rhom(\vx) \mathrm{d}\vx = 
    (\Ne)^p \EX_{\rho_\phi} [ f(\vx, \rho_\phi,\nabla\rho_\phi) ],
    \label{eqn:expec_rho}
\end{eqnarray}
where $(\Ne)^p$ is the constant factor related to the number of electrons, and $f(\vx,\rho_\phi,\nabla\rho_\phi)$ is the integrand of the functional $F[\rhom]$. All functionals are estimated with Monte Carlo,
where the samples are drawn from $\rhozero(\vz)$ and transformed by a CNF, $\vx = T_\phi(\vz) := \vz + \int_{t_{0}}^{T} g_{\phi}(\vz(t),t) \mathrm{d}t$.

Similar to any variational protocol in many-body physics, the minimization of total energy can be done through gradient-based algorithms using automatic differentiation (AD) ~\citep{arrazola2023differentiable,kasim2021dqmc,tamayo2018hfad,vargashdz2023huxel}. AD for OF-DFT with grids was proposed by \citet{tan2023automatic}.
For all the results presented here, $\rhozero(\vz)$ is a Gaussian distribution with an identity covariance matrix.
We found the $\texttt{RMSProp}$ \citep{hinton2012rmsprop} algorithm with a learning rate of $3\times10^{-4}$ to be the most appropriate optimizer. Code was developed using \texttt{JAX} Ecosystem~\citep{jax2018github,flax2020github}.\vspace{-10pt}
The code is available in the following \href{https://github.com/RodrigoAVargasHdz/ofdft_normalizing-flow}{GitHub Repository}. 

\vspace{-0.1cm}
\section{Results}
\vspace{-.2cm}
In this section, we present the results for a one-dimensional model for diatomic models, specifically \texttt{LiH}, and the simulations for hydrogen and water molecules. 

\vspace{-.2cm}
\subsection{1-D model for diatomic molecules}
\vspace{-.2cm}
Based on \citep{snyder2013orbital} work, we considered a one-dimensional model for diatomic molecules where the total energy functional is defined as,
\begin{equation}
    E[\rhom] = T[\rhom] + V_{\text{H}}[\rhom] +  V_{\text{e-N}}[\rhom]  + E_{\text{X}}[\rhom]. 
    \label{eqn:energy_func}
\end{equation}
The total kinetic energy is estimated by the sum of the Thomas-Fermi (TF) (Eq. \ref{eqn:thomas_fermi}) and  Weizs\"{a}cker (Eq. \ref{eqn:weizsacker}) functionals~\citep{gpaar1980_DFT}; $T[\rhom] = T_{\text{TF}}[\rhom] + T_{\text{W}}[\rhom]$. 
\noindent\begin{minipage}{0.46\textwidth}
\begin{footnotesize}
\begin{equation}
    T_{\text{TF}}[\rhom] = \frac{3}{10}(3\pi^2)^{\frac{2}{3}} \int \left(\rhom(\vx) \right)^{5/3} \mathrm{d}\vx \label{eqn:thomas_fermi}
\end{equation} 
\end{footnotesize}
    \end{minipage}%
    \begin{minipage}{0.125\textwidth}\centering
    and 
    \end{minipage}%
    \begin{minipage}{0.4\textwidth}
\begin{footnotesize}
\begin{equation}
    T_{\text{W}}[\rhom] = \frac{\lambda}{8} \int \frac{(\nabla \rhom(\vx))^2}{\rhom} \mathrm{d}\vx, \label{eqn:weizsacker}
\end{equation}
\end{footnotesize}
    \end{minipage}
where the phenomenological parameter $\lambda$ was set to 0.2~\citep{chan2001thomas}. 
For one-dimensional systems $T_{\text{TF}}$ is $T_{\text{TF}}[\rhom] = \pi^2/24 \int \left(\rhom(x) \right)^{3} \mathrm{d}x$ \citep{snyder2013orbital} and $T_{\text{W}}$ and $E_{\text{X}}$ have the same analytic form as in Eqs. \ref{eqn:weizsacker} and \ref{eqn:exchange_dirac} respectively.
We rewrite $T_{\text{W}}[\rhom]$ in terms of the score function;  $T_{\text{W}}[\rhom] = \frac{\lambda}{8} \int  \left(\nabla \log \rhom(x) \right)^2  \rhom(x) \mathrm{d}x$, and in the CNF framework it can be computed by solving,
\begin{equation}
    \tfrac{\mathrm{d}}{\mathrm{d} t} \nabla \log \rho_\phi = 
    -\nabla^2 g_{\phi}
    - \left(\nabla \log \rho_\phi\right)^T \left(\nabla g_{\phi}(\mathbf{z}(t), t) \right),
    \label{eqn:score_of_cnf}
\end{equation}
where $\nabla^2$ is the Laplacian operator. This allows us to make use of memory-efficient gradients \citep{chen2023odescore} for optimizing $T_{\text{W}}[\rhom]$.

\begin{figure}[!ht]
    \centering
    \begin{subfigure}[b]{0.33\linewidth}
        \includegraphics[width=\linewidth]{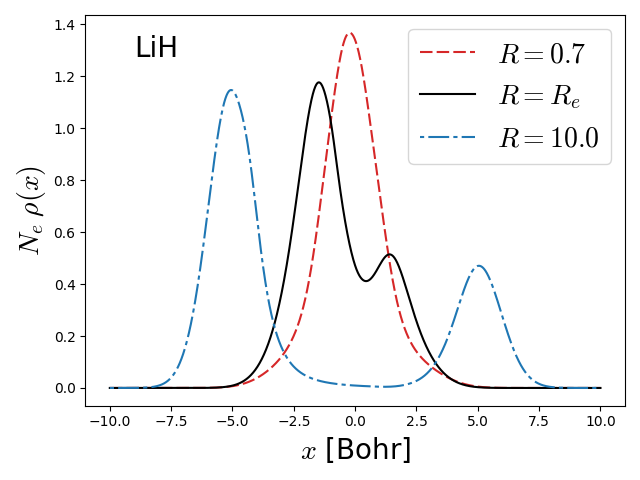}
        \vspace{-1.3em}
        \caption{\texttt{LiH} density}
        \label{fig:1d_lih_a}
    \end{subfigure}%
    \begin{subfigure}[b]{0.33\linewidth}
        \includegraphics[width=\linewidth]{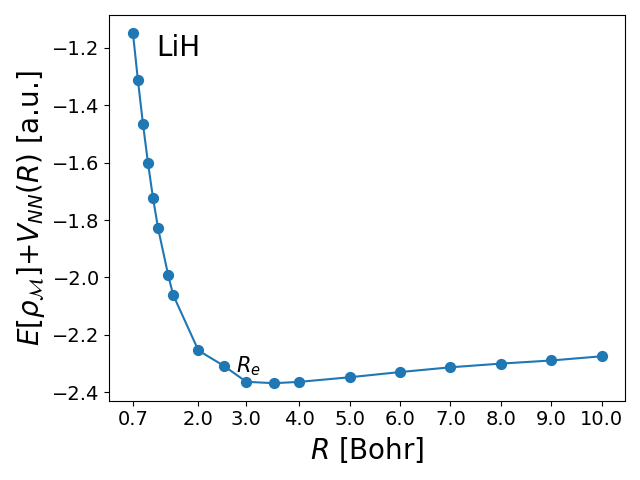}
        \vspace{-1.3em}
        \caption{Potential energy surface}
        \label{fig:1d_lih_b}
    \end{subfigure}%
    \begin{subfigure}[b]{0.33\linewidth}
        \includegraphics[width=\linewidth]{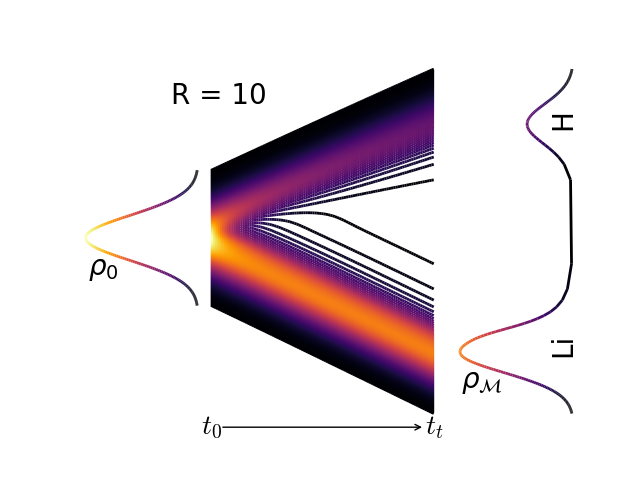}
        \vspace{-1.3em}
        \caption{$T_\phi(x(t))$}
        \label{fig:1d_lih_c}
    \end{subfigure}%
    \caption{(a) Ground state electronic density of \texttt{LiH} for various inter-atomic distances $R$. (b) The value of the potential energy as a function of $R$ is computed with $E[\rhom] + V_{\text{NN}}(R)$\footnotemark. (c) The  change of $x(t)$ and $\rho_\phi(x(t))$ through the transformation $T_\phi$ (Eq. \ref{eqn:cnf_evol}). For all simulations, $\rhozero$ is a Gaussian distribution with $\sigma=1$.}
    \label{fig:LiH}
    \vspace{-0.5cm}
\end{figure}

The Hartree ($V_{\text{H}}[\rhom]$) potential and the external potential ($V_{\text{e-N}}[\rhom]$) functionals both are defined by a soft version \citep{snyder2013orbital}, \vspace{-0.2cm}
\begin{footnotesize}
\begin{align}
    V_{\text{H}}[\rhom] &= \int \int v_{\text{H}}(x) \rhom(x)\rhom(x')\mathrm{d}x \mathrm{d}x' = \int \int \tfrac{ \rhom(x)\rhom(x')}{\sqrt{1 + |x - x'|^2}} \mathrm{d}x \mathrm{d}x',  \label{eqn:hartree}   \\
    V_{\text{e-N}}[\rhom] &= \int v_{\text{e-N}}(x) \rhom(x) \mathrm{d}x = -\int  \left  ( \tfrac{Z_\alpha}{\sqrt{1 + | x - R/2 |^2}} + \tfrac{Z_\beta}{\sqrt{1 + | x + R/2 |^2}} \right )\rhom(x) \mathrm{d}x.
    \label{eqn:ext_pot}
\end{align}
\end{footnotesize}
We only consider the Dirac exchange functional \citep{gpaar1980_DFT}, \vspace{-0.1cm}
\begin{footnotesize}
\begin{equation}
    E_{\text{X}}[\rhom] = -\frac{3}{4} \left( \frac{3}{\pi} \right)^{1/3} \int \rhom^{4/3}(\vx) \ \ \mathrm{d}\vx.
    \label{eqn:exchange_dirac}
\end{equation}
\end{footnotesize}
We use the Lithium hydride molecule (\texttt{LiH}) as an example of diatomic molecule modeling in one dimension. For \texttt{LiH}, we utilize\footnotemark: $Z_{\alpha} = 3$, $Z_{\beta} = 1$, $N_e = 2$ (valence electrons), and various nuclear distances $R$. In Fig. \ref{fig:1d_lih_a}, we present $\rhom$ for various $R$ values, each one corresponding to the training of independent CNF models using only 2,000 iterations. Additionally, we display the potential energy surface curve in Fig. \ref{fig:1d_lih_b}, with the equilibrium bond distance found at $R_e = 2.95$ Bohr \citep{snyder2013orbital}. By exploring different $R$ values, we validate the capability of our CNF ansatz for $\rhom$ to parameterize diverse chemical scenarios, ranging from strong nuclear interactions ($R<R_e$) to bond-breaking ($R \gg R_e$). The latter case is found to be challenging for existing DFT methodologies \citep{mori2009discontinuous}. From Fig. \ref{fig:1d_lih_a}, it is evident that $\rhom$ does not distribute electron density along the inter-atomic axis. Fig. \ref{fig:1d_lih_c} we illustrate the transformation $\rho_0$ to $\rhom$ through the ODE (Eq. \ref{eqn:cnf_evol}). Notably, for $R=10$, $\rhom$ exhibits a bimodal distribution, with a higher concentration of electron density nearer to the \texttt{Li} nucleus.
\footnotetext{$Z_i$ is the atomic number of atom $i$.}
\footnotetext{$V_{\text{NN}}(R) = Z_{\alpha} Z_{\beta} / \sqrt{1 + R^2}$}

\vspace{-0.2cm}
\subsection{3-D simulations}
\vspace{-0.3cm}
To demonstrate the capability to use CNF for real-space simulations, we considered the optimization of $\texttt{H}_{2}$ and $\texttt{H}_{2}\texttt{O}$. For both chemical systems, we considered the same $E[\rhom]$ where $v_{\text{e-N}}(\vx)$; $v_{\text{e-N}}(\vx) = -\sum_i \frac{Z_i}{\|\vx - \mathbf{R}_i\|}$. 
For these simulations, no soft approximation was considered for the $V_{H}$ and $V_{\text{e-N}}$ functionals.

Fig.\ref{fig:MEP} illustrates the change in $\rhom$ through the optimization procedure for \texttt{H}$_2$ and \texttt{H}$_2$\texttt{O}. 
To further illustrate the change of $\rhom$ from $\rhozero$ by minimizing $E[\rhom]$, we computed the molecular electrostatic potential (MEP); $V_{\text{MEP}}(\vx) = v_{\text{e-N}}(\vx) - \int \rhom(\vx')/\|\vx - \vx'\| \mathrm{d}\vx'$.
For both systems, $\rhozero$ is a multivariate Gaussian distribution located at zero, and the last layer of $g_{\phi}$ was initialized to zero. We compare our results with Hartree-Fock (HF) with STO-3G basis set ($\hrhom$). We monitored the value of $\int \rhom(\vx) \mathrm{d}\vx$ using Becke's integration Scheme \citep{becke1988multicenter}; however, we found no significant difference w.r.t. $\Ne$, $\left |\int \rhom(\vx) \mathrm{d}\vx - \Ne \right | < 10^{-4}$. 

\begin{figure}
    \centering
    \begin{subfigure}[b]{0.33\linewidth}
        \includegraphics[width=\linewidth]{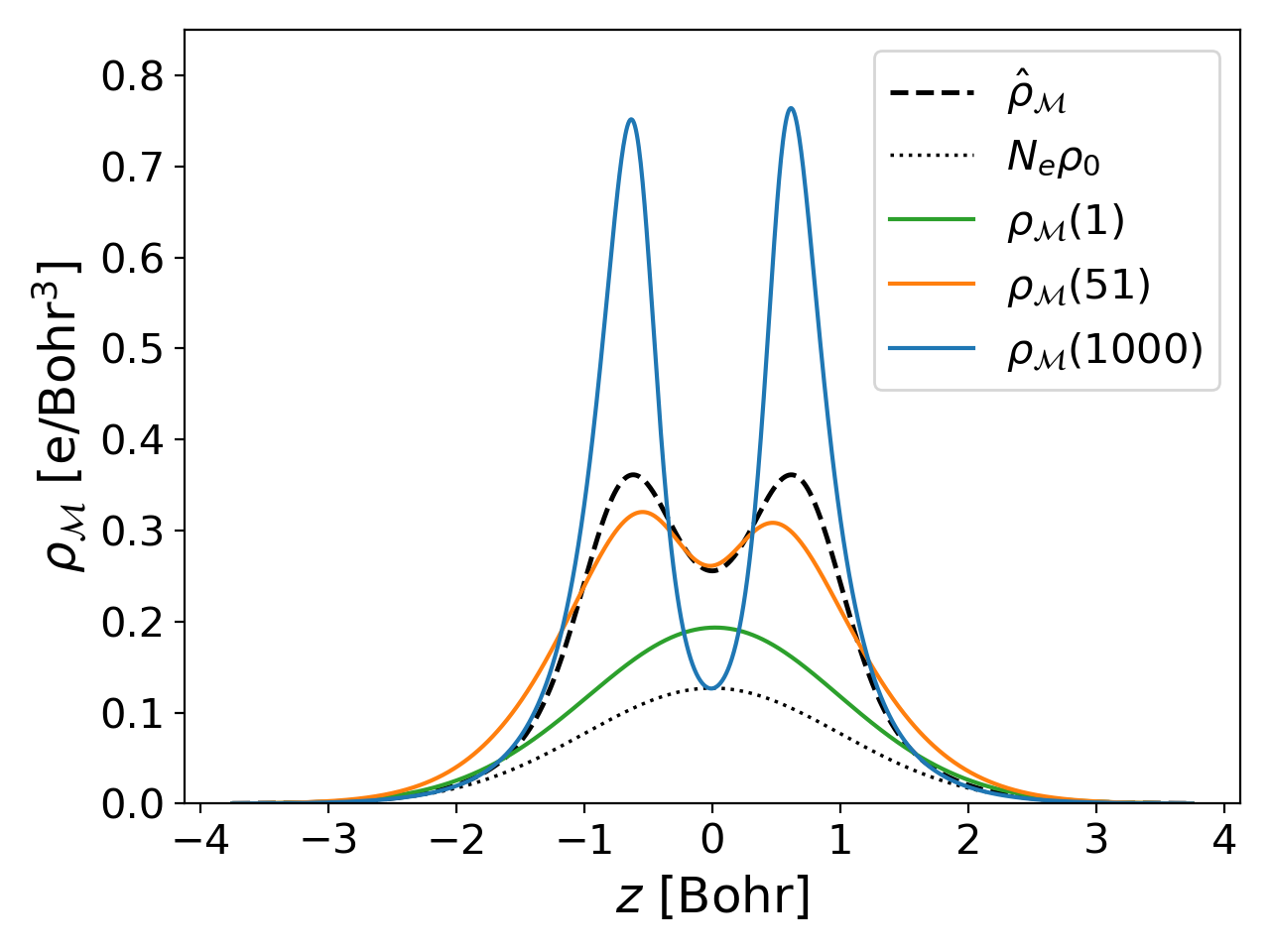}
        \vspace{-1.3em}
        \caption{\texttt{H}$_2$ density}
        \label{fig:3d_h2_a}
    \end{subfigure}%
    \begin{subfigure}[b]{0.33\linewidth}
        \includegraphics[width=\linewidth]{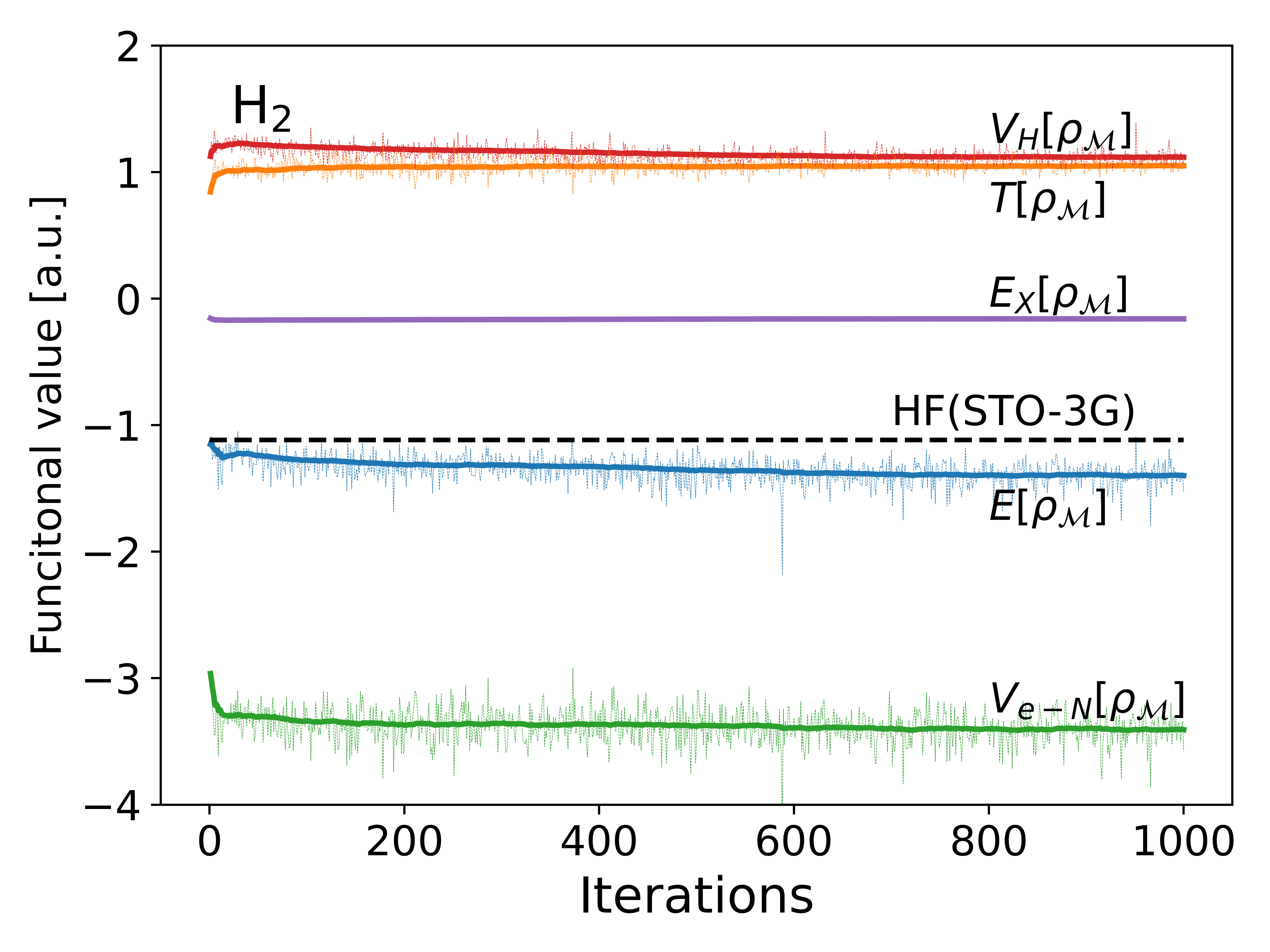}
        \vspace{-1.3em}
        \caption{\texttt{H}$_2$ training}
        \label{fig:3d_h2_b}
    \end{subfigure}%
    \begin{subfigure}[b]{0.33\linewidth}
        \includegraphics[width=\linewidth]{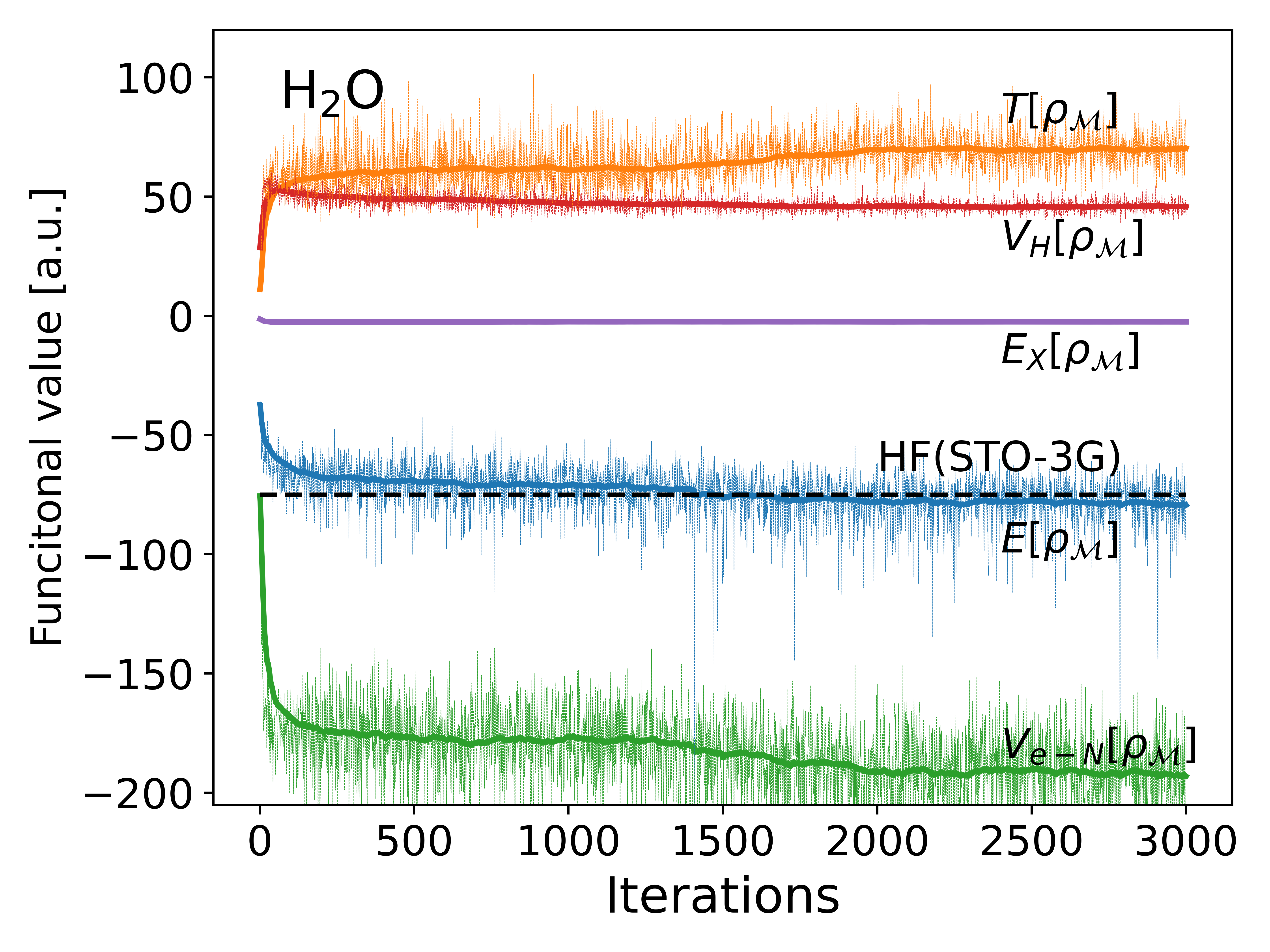}
        \vspace{-1.3em}
        \caption{\texttt{H}$_2$\texttt{O} training}
        \label{fig:3d_h2o}
    \end{subfigure}%
    \caption{(a) A cross-section of $\rhom(\texttt{i})$ at various iterations $\texttt{i}$. (b, c) The value of each density functional throughout the optimization, all values computed using an exponential moving average. }
    \label{fig:cross_section}
\end{figure}

For the hydrogen molecule, the CNF model took 50 iterations to transport the density mass towards the \texttt{H} nuclei (Fig. \ref{fig:MEP}), $R=1.4$ Bohr. After 1,000 iterations we observe no significant difference in $E[\rhom]$ (Fig. \ref{fig:3d_h2_b}) and we observe a higher density around the nuclei compared to $\hrhom$, Fig. \ref{fig:3d_h2_a}. 

For the optimization of the water molecule (Fig. \ref{fig:3d_h2o}), our procedure was able to describe the two electron pairs, characteristic of this chemical system. We can also observe the change on $\rhozero$, a fully radial symmetric density, to a density with a higher electron density closer to the $\texttt{O}$ atom. We set the maximum number of iterations to 3,000 with a 512 batch size, however, a $\rhozero$ closer to $\rhom$ could reduce the number of iterations.\looseness=-1



\vspace{-0.4cm}
\section{Summary}
\vspace{-0.3cm}
We have introduced a novel numerical procedure for parametrizing the ground-state energy of molecular densities using normalizing flows within the OF-DFT framework. This approach optimizes $\rhom$ variationally by minimizing the total energy $E[\rhom]$, marking it as the first constraint-free method that ensures both normalization and positivity (Eq. \ref{eqn:ofdft_min}). Our study encompasses two types of simulations: one involving one-dimensional diatomic molecules and another involving a comprehensive real-space simulation of the \texttt{H}$_2$ and \texttt{H}$_2$\texttt{O} systems.



\bibliographystyle{plainnat-sortby-first}
\bibliography{library}


\end{document}